\documentclass[aps,pra,reprint,amsmath,amssymb,superscriptaddress,floatfix,showpacs,showkeys]{revtex4-1} 
\usepackage{hyperref}
\hypersetup{
   unicode=true,          
   plainpages=false,
   colorlinks=true,       
   citecolor=blue        
}
\usepackage{graphics} 
\usepackage{graphicx} 
\usepackage{subfigure}
\usepackage[normalem]{ulem}
\usepackage{url}
\usepackage[utf8]{inputenc}
\usepackage{braket}
\newcommand{\rme}{\mathrm{e}}
\newcommand{\rmi}{\mathrm{i}}
\newcommand{\x}{\mathrm{x}}
\newcommand{\y}{\mathrm{y}}
\renewcommand{\d}{\ensuremath{\mathrm{d}}}

\newcommand{\field}[1]{\mathbb{#1}} 
\usepackage{xcolor}

\begin{document} 
\author{L. A. Toikka}
\affiliation{Dodd-Walls Centre for Photonic and Quantum Technologies, Centre for Theoretical Chemistry and Physics, and New Zealand Institute for
Advanced Study, Massey University, Private Bag 102904 NSMC, Auckland 0745, New
Zealand}
\title{Boson-vortex duality in compressible spin-orbit coupled BECs}
\date{\today}
\begin{abstract}
Using a (1+2)-dimensional boson-vortex duality between non-linear electrodynamics and a two-component compressible Bose-Einstein condensate (BEC) with spin-orbit (SO) coupling, we obtain generalised versions of the hydrodynamic continuity and Euler equations where the phase defect and non-defect degrees of freedom enter separately. We obtain the generalised Magnus force on vortices under SO coupling, and associate the linear confinement of vortices due to SO coupling with instanton fluctuations of the dual theory. \looseness=-1
\end{abstract}
\maketitle

\textit{Introduction:} The recent ground-breaking experimental developments of spin-orbit (SO) coupling in ultra-cold atomic gases continue to highlight the importance of these highly controllable systems as emulators of condensed matter. For bosons~\cite{Lin2011,Galitski2013}, this was achieved following the earlier synthetic creation of an artificial magnetic field~\cite{Lin09}, which makes it possible to create stationary vortices in a non-rotating condensate, but the impact of SO coupling goes deeper. In particular, the dynamics of quantised vortices is affected in a non-trivial way as a result of an additional contribution to the vortex force by the SO and Rabi couplings, which we derive here. We present a hydrodynamic description of SO coupled BECs that directly includes dynamics of the vortex degrees of freedom. 

Typical superfluid hydrodynamics is formulated using the density $n$ and velocity $\vec{v} = \frac{\hbar}{m} \vec{\nabla} S$, where $S$ is the superfluid phase, through the Madelung transformation $\psi = \sqrt{n} \rme^{\rmi S}$. Here $\psi$ satisfies the Gross-Pitaevskii equation (GPE). However, $S$ is multi-valued, which creates often overlooked but important problems with the normal rules of calculus~\cite{kleinert2008multivalued}. This is especially manifest in the presence of a quantised vortex, which creates a logarithmic branch cut in the phase of the incompressible sector. To this end, various gauge fields have been introduced to address this problem~\cite{Klein2014195,Kozhevnikov2015122,PhysRevA.94.063633,PhysRevA.94.063623}, but it remains unclear what happens in the presence of a two-component compressible BEC under SO coupling.

To answer this question, we establish a boson-vortex duality that allows us to map the superfluid dynamics onto non-linear electrodynamics, i.e. a $U(1)$ gauge theory. In the particle-vortex dualities~\cite{PhysRevX.6.031043,doi:10.1142/S0217979290000206,doi:10.1142/S0217979291001061}, the underlying phonon or the superfluid component enters as a gauge potential $A_\mu$, while vortices represented by phase defects emerge as fundamental bosonic fields i.e. point-like particles carrying charge with respect to $A_\mu$. The key idea is that the defect and non-defect degrees of freedom are clearly separated, which makes it most natural to study the dynamics of vortices (and other phase defects). We formulate a partition function from which we can obtain important physical results, in fact all the quantum properties, by suitable differentiation provided that the actual path integral can be evaluated. In this case vortices are bosonic particles, and the quantum description involves an integration over all possible worldlines of the vortex degrees of freedom, weighted by the Lagrangian that we derive.

In general, SO coupling links a particle's momentum to its spin state. In the BEC experiments, the two pseudo-spin states, formed by two hyperfine states that have been split using an external Zeeman field, are coupled by two polarized Raman beams. The Raman laser wave number difference $k_0$ and Rabi frequency $\Omega$ (as well as the Zeeman field $\delta$) are all under experimental control, and they are related to the dipole coupling between the hyperfine states by $(\hbar \Omega/2)\rme^{2\rmi k_0 x} = \bra{\uparrow} \textbf{d} \cdot \textbf{E} \ket{\downarrow}$. In the limit as $k_0 \to 0$, the Rabi frequency will act as a spatially homogeneous tunneling coupling between the states, reducing the problem to two coherently tunnel-coupled condensates~\cite{PhysRevA.73.013627,2016arXiv160903966C}. The synthetic gauge field thus obtained, $-k_0 \sigma_\mathrm{z}$, consists of only one component of the synthetic vector potential, and is therefore Abelian. 

The key result of this work are three equations, characterising the hydrodynamics of the SO coupled BEC. The first is the continuity equation, the second one represents the generalised Euler equation, and the third one enforces the quantization of circulation of vortices. In addition, we show how SO coupling changes the net force on vortices.

\textit{Boson-vortex duality for spin-orbit coupling:} We start by formulating the zero-temperature hydrodynamical theory for a two-component (pseudo-spin $1/2$) SO coupled BEC in terms of a partition function. We focus on two spatial dimensions $x$ and $y$. Throughout this work we adopt the convention that two-dimensional spatial vectors are denoted by an arrow, while three-vectors are denoted by bold symbols. The supefluid velocity is given by $\textbf{v}^{(l)} = \frac{\hbar}{m} (\frac{1}{c}\dot{\tilde{S}}_l, \vec{\nabla} \tilde{S}_l)^\mathrm{T} \equiv (v_1^{(l)},\vec{v}^{(l)})^\mathrm{T} $, where $l = 1,2$ is the component index, $\tilde{S}_l$ ($S_l$) is the phase of component $l$ in the presence (absence) of SO coupling, and $m$ is the mass of the bosons. Here $c = \sqrt{n_\infty g / m} $ is the speed of sound of a uniform condensate of density $n_\infty$, where $n_\infty$ is the density at infinity far away from vortices. We define the 3-gradient $\nabla = \partial_\mu = (\frac{1}{c}\partial_t, \partial_x, \partial_y)^\mathrm{T}$.

The partition function (full quantum theory) can be written as $\mathcal{Z} = \int \mathcal{D}\chi \mathcal{D}\chi^\dagger\, \rme^{\frac{\rmi}{\hbar} \mathcal{S}[\chi, \chi^\dagger] }$~\cite{popov2001functional}, where the action is given by $\mathcal{S}[\chi, \chi^\dagger] = \int \d^3 x\, \mathcal{L}[\chi, \chi^\dagger]$ with the Lagrangian density
\begin{equation}
\mathcal{L}[\chi, \chi^\dagger] = \frac{\rmi \hbar}{2} \left( \chi^\dagger \partial_t \chi - (\partial_t \chi^\dagger) \chi \right) - E[\chi, \chi^\dagger].
\end{equation}
Here $\int \d^3 x \equiv \int c\d t \d x \d y$, and $\chi$ is a two-component spinor. The energy is given by the Gross-Pitaevskii energy functional
\begin{equation}
\label{eqn:E}
E\left[\chi, \chi^\dagger \right] = \chi^\dagger h_0 \chi + \frac{g}{2}\sum_l |\chi_l|^4 + g_{12} |\chi_1|^2|\chi_2|^2,
\end{equation}
where the single-particle Hamiltonian $h_0$ is given by~\cite{Lin2011}
\begin{equation}
\label{eqn:h0}
h_0 = \frac{\hbar^2}{2m}\left(-\rmi \vec{\nabla} \sigma_0 + k_0 \hat{\textbf{x}} \sigma_\mathrm{z} \right)^2 + \frac{\hbar \delta}{2} \sigma_\mathrm{z} + \frac{\hbar \Omega}{2} \sigma_\mathrm{x} + V_\mathrm{tr} \sigma_0.
\end{equation}
Here $\sigma_0$ is the $2\times 2$ identity matrix, $\sigma_i$ with $i = x,y,z$ are the Pauli matrices, $g$ and $g_{12}$ are Gross-Pitaevskii coupling constants, $k_0$ and $\Omega$ are the wavenumber difference and Rabi frequency of the Raman lasers inducing the SO coupling respectively, $\delta$ is a Zeeman term due to an external magnetic field, and $V_\mathrm{tr}$ is the external trapping, which we do not specify here. 

We consider the following general form for the spinor:
\begin{equation}
\label{eqn:chi}
\chi =   \begin{pmatrix}
\sqrt{n_1}  \rme^{\rmi S_1}\\
\sqrt{n_2}  \rme^{\rmi S_2}
\end{pmatrix},
\end{equation}
where $n_l$ is the density of component $l$. Then $\mathcal{L} = T + T_\mathrm{n} - V_\mathrm{n}$, where
\begin{subequations}
\begin{align}
\label{eqn:Lagformdefs-a}
V_\mathrm{n} &= \hbar \Omega \sqrt{n_1 n_2}\cos{\left( \tilde{S}_1 - \tilde{S}_2  - 2k_0 x\right)},\\
\label{eqn:Lagformdefs-b}
T &=  -\hbar \left(n_1  \dot{\tilde{S}}_1 + n_2  \dot{\tilde{S}}_2 \right) - \frac{\hbar^2 n_1  |\vec{\nabla} \tilde{S}_1|^2}{2m}     - \frac{\hbar^2 n_2 |\vec{\nabla} \tilde{S}_2|^2}{2m}  ,\\
\label{eqn:Lagformdefs-c}
T_\mathrm{n} &= -\frac{\hbar \delta}{2} \left( n_1 -  n_2 \right)    - \frac{\hbar^2}{2m} \left(  \frac{ |\vec{\nabla} n_1|^2 }{4n_1} +  \frac{ |\vec{\nabla} n_2|^2 }{4n_2} \right)  \\
&\notag  \qquad  - V_{\mathrm{tr}} \left( n_1 +  n_2 \right) - \frac{g}{2}\left( n_1^2 +  n_2^2 \right) - g_{12}   n_1 n_2,
\end{align}
\end{subequations}
where we have defined $\tilde{S}_l \equiv S_l \pm k_0 x$ with $+,-$ for $l=1,2$. The canonical momentum is obtained as $p_0^{(l)} = \partial \mathcal{L}/\partial \left(\frac{1}{c} \dot{\tilde{S}}_l \right) = -\hbar cn_l$. We now introduce a Hubbard-Stratonovich field $\vec{p}^{(l)}$ so that the path integral has to be performed over $\textbf{p}^{(l)} = (p_0^{(l)}, \vec{p}^{(l)})^\mathrm{T}$ and $\tilde{S}_l$:
\begin{equation}
\label{eqn:LagZL-H-u}
\begin{split}
\mathcal{Z} &=    
\int \mathcal{D}\tilde{S}_1 \mathcal{D}\tilde{S}_2 \mathcal{D}\textbf{p}^{(1)}\mathcal{D}\textbf{p}^{(2)} \exp \left\lbrace  \frac{\rmi}{\hbar} \int \d^3 x \left[T_\mathrm{B} - V_\mathrm{B} \right.\right. \\
 &\left. \left.  +\sum_{l=1,2}{\left( \frac{1}{c} \dot{\tilde{S}}_l\, p_0^{(l)} \,\substack{+\\(-)}  \vec{\nabla}\tilde{S}_l \cdot \vec{p}^{(l)} -  \frac{cm \left(\vec{p}^{(l)}\right)^2}{2 \hbar p_0^{(l)}}   \right)} \right] \right\rbrace,
\end{split}
\end{equation}
where $T_\mathrm{B}$ and $V_\mathrm{B}$ are obtained from $T_\mathrm{n}$ and $V_\mathrm{n}$ respectively by expressing $n_l = -p_0^{(l)}/(\hbar c)$. Equation~\eqref{eqn:LagZL-H-u} is the full quantum theory for a SO coupled BEC in terms of the Gross-Pitaevskii variables, adding all the quantum fluctuations around the mean field.

The physical meaning of $\vec{p}^{(l)}$ becomes clear by evaluating its Euler-Lagrange equation of motion using Eq.~\eqref{eqn:LagZL-H-u}: $\vec{p}^{(l)} = -\frac{\hbar^2}{m}  n_l \vec{\nabla} \tilde{S}_l \equiv -\hbar n_l \vec{v}^{(l)} \equiv -\hbar \vec{j}^{(l)}$; it is the superfluid current. Together they form the 3-vector $\textbf{j}^{(l)} = (c n_l, n_l \frac{\hbar }{m}\vec{\nabla} \tilde{S}_l)^\mathrm{T}$. We note that the sign of the term $\vec{\nabla}\tilde{S}_l \cdot \vec{p}^{(l)}$ can be chosen arbitrarily, and therefore has no effect on $\mathcal{Z}$. We take the sign to be +.

In general, the non-linear term $V_\mathrm{B}$ prevents a direct integration over the fields $\tilde{S}_{1,2}$. On the other hand, in the phase-locked regime of $S_1 = S_2$~\cite{PhysRevA.86.063621}, for example, or taking a vanishing Rabi frequency $\Omega = 0$ an exact integration over $\tilde{S}_{1,2}$ would be possible leading to the continuity constraint $\nabla \cdot \textbf{j}^{(l)} = 0$. To see this, we integrate by parts so that the phase appears only in the form $- \tilde{S}_l \nabla \cdot \textbf{p}^{(l)} = \hbar \tilde{S}_l \nabla \cdot \textbf{j}^{(l)} $ in the action~\eqref{eqn:LagZL-H-u}. The constraint is resolved by writing $\textbf{j}^{(l)} \propto \nabla \times \textbf{A}^{(l)}$, where $\textbf{A}^{(l)} = (a_0^{(l)},-\vec{a} ^{(l)})^\mathrm{T} = (a_0^{(l)},-a_\x ^{(l)},-a_\y ^{(l)})^\mathrm{T}$ is unconstrained. The functional integral over $\textbf{p}^{(l)}$ is then replaced by a functional integral over $\textbf{A}^{(l)}$, arriving at the Popov formulation of non-linear electromagnetism~\cite{popov2001functional}.

Instead, we replace the integration over $\tilde{S}_{1,2}$ by the corresponding classical action $\mathcal{S}_\mathrm{cl}$. This means that we substitute $\tilde{S}_{l} \to \tilde{S}_l^\mathrm{cl}$, where $\tilde{S}_l^\mathrm{cl}$ solves the Euler-Lagrange equation for $\tilde{S}_{l}$, in the action~\eqref{eqn:LagZL-H-u}. This procedure is exact only for quadratic Lagrangians~\footnote{$\int_{x_1}^{x_2}\mathcal{D}x(t) \exp{\left(\frac{\rmi}{\hbar}\int_{t_1}^{t_2} \mathcal{L}(x,\dot{x}) \d t  \right)} \propto \exp{\left(\frac{\rmi}{\hbar} \mathcal{S}_\mathrm{cl}(x_1,x_2)\right)}$, where $\mathcal{L}(x,\dot{x})$ denotes any Lagrangian that is at most a quadratic form in $x$ and $\dot{x}$, and $\mathcal{S}_\mathrm{cl}(x_1,x_2) = \int_{t_1}^{t_2} \mathcal{L}(x_\mathrm{cl},\dot{x}_\mathrm{cl}) \d t$ is the corresponding classical action, evaluated along the classical trajectory $x_\mathrm{cl}$ with the boundary conditions $x_\mathrm{cl}(t_i) = x_i$ for $i = 1,2$~\cite{1993hep.th....2053G}.}, in particular if $\Omega = 0$. One can then expect the approximation to be accurate at least for small $\Omega$. In our case, the constraint (i.e. Euler-Lagrange equation for $\tilde{S}_l)$ amounts to the continuity equation, which under SO coupling thus becomes 
\begin{equation}
\label{eqn:conteq}
\nabla \cdot \textbf{j}^{(l)} = \pm \Omega \sqrt{n_1 n_2}\sin{\left( \zeta\right)},
\end{equation}
where the $+$ $(-)$ corresponds to $l = 1$ ($l=2$), and $\zeta \equiv \tilde{S}_1^\mathrm{cl} - \tilde{S}_2^\mathrm{cl}  - 2k_0 x$. This time, we resolve the continuity constraint by writing a more general Helmholtz decomposition 
$\textbf{j}^{(l)} = - \nabla \times \textbf{A}^{(l)} + \nabla \phi_l$, 
where the curl-free longitudinal component is given by
\begin{equation}
\label{eqn:phi}
\phi_l(\textbf{r})  = \pm \Omega  \int \d^3 x^\prime \, \left[\sqrt{n_1 n_2}\sin{\left( \zeta\right)}\right](\textbf{r}^\prime) G(\textbf{r},\textbf{r}^\prime),
\end{equation}
where the $+$ $(-)$ corresponds to $l = 1$ ($l=2$) and $G$ is the Green's function. In the presence of SO coupling, the resolution of the constraint amounts to 
\begin{subequations}
\label{eqn:EB}
\begin{align}
cn_l &=  B^{(l)} + \frac{1}{c}\partial_t \phi_l, \\ 
j_\x^{(l)} &= E_\y^{(l)} +  \partial_x \phi_l   ,\\
j_\y^{(l)}  &=  -E_\x^{(l)} + \partial_y \phi_l ,
\end{align}
\end{subequations}
where we have defined an `electric field' $\vec{E}^{(l)} = -\frac{1}{c}\partial_t \vec{a}^{(l)} - \vec{\nabla} a_0^{(l)} = -\hat{\epsilon} \left(\vec{j}^{(l)} - \vec{\nabla} \phi_{l} \right)$, and a `magnetic flux' $B^{(l)} = \vec{\nabla} \times \vec{a}^{(l)}$. The operator $\hat{\epsilon} = \rmi \sigma_\y$ is the fully anti-symmetric Levi-Civita tensor of rank 2, corresponding to a rotation in the $xy$-plane by the angle $-\pi/2$.

We note that $\phi_l$ is not a new degree of freedom, but rather a short-hand for the integral~\eqref{eqn:phi}, fixed by the divergence of $\textbf{j}^{(l)}$ (Eq.~\eqref{eqn:conteq}). However, the curl of $\textbf{j}^{(l)}$ is not given by the Euler-Lagrange equations for $\tilde{S}_l$ alone, which in fact needs all the remaining equations of motion. The aim of the rest of this work is to find equations for $\vec{E}^{(l)}$ and $B^{(l)}$ such that Eqs.~\eqref{eqn:EB} are satisfied, but where vortices appear explicitly.

The $U(1)$ gauge theory for the fields $\textbf{A}$ can be coupled to charges. If $\textbf{J}$ denotes the $U(1)$ charge 3-current, the minimal coupling procedure $-\rmi \hbar \nabla \to -\rmi \hbar \nabla - q \textbf{A}$ that couples the charges to the gauge fields amounts to having the term $\textbf{A} \cdot \textbf{J}$ in the coupled Lagrangian. This is an example of the celebrated boson-vortex duality~\cite{doi:10.1142/S0217979291001061,doi:10.1142/S0217979290000206,PhysRevX.6.031043,Wen2004} in which the quantised point charges in the $U(1)$ gauge theory correspond to extended superfluid vortices. Assuming the presence of regular vortices in the two components, we define the vortex 3-current 
\begin{equation}
\label{eqn:vortexcurrent}
\begin{split}
\textbf{J}^{(l)} = \begin{pmatrix}
J_0^{(l)} \\
\vec{J}^{(l)}
\end{pmatrix}
=
\begin{pmatrix}
\sum_{j} q_{j}^{(l)} \delta \left[\vec{r} - \vec{R}^{(l)}_{j}(t)\right] \\
\sum_{j} q_{j}^{(l)} \frac{1}{c} \partial_t \vec{R}^{(l)}_{j}(t) \delta \left[\vec{r} - \vec{R}^{(l)}_{j}(t)\right]
\end{pmatrix},
\end{split}
\end{equation}
where we have taken $m_l$ vortices in component $l$, of circulation $q_{j}^{(l)} \in \field{Z}$, located at $\vec{R}_{j}^{(l)}(t)$ with $j = 1, 2, \ldots, m_l$. It satisfies the continuity equation $\nabla \cdot \textbf{J}^{(l)} = 0$. We must then integrate over all possible vortex positions in the functional integral. We ignore the complication arising from the fact that strictly speaking the functional integral over $\textbf{A}^{(l)}$ is coupled to the vortex positions because the condensate density has to vanish at the vortex cores. In other words, we take the functional integration over $\textbf{A}^{(l)}$ to be independent from the functional integration over all the vortex positions.

Thus, replacing the functional integral over $\tilde{S}_{1,2}$ by the corresponding classical action $\mathcal{S}_\mathrm{cl}$ in Eq.~\eqref{eqn:LagZL-H-u}, shifting the integration variable $\textbf{p}^{(l)}$ by $\nabla \phi_l$, and finally writing the path integral in terms of the unconstrained fields $\textbf{A}^{(l)}$, we obtain the partition function
\begin{equation}
\label{eqn:Z}
\begin{split}
\mathcal{Z} &= \left[\prod_{j}\prod_{k} \int \mathcal{D}\vec{R}^{(1)}_{j}(t)\mathcal{D}\vec{R}^{(2)}_{k}(t) \right]\int \mathcal{D}\textbf{A}^{(1)}\mathcal{D}\textbf{A}^{(2)} \rme^{ \frac{\rmi}{\hbar} \mathcal{S}},
\end{split}
\end{equation}
where the action $\mathcal{S}$ is given by
\begin{equation}
\label{eqn:SOC_action}
\begin{split}
\mathcal{S}  &=  \int \d^3 x \left[ -\frac{\hbar \delta}{2} \left( \frac{B^{(1)}}{c} -  \frac{B^{(2)}}{c} \right) - \frac{\hbar \Omega}{c}  \sqrt{B^{(1)} B^{(2)} }\cos{\left( \zeta \right)} \right. \\
&  \qquad \qquad - \frac{g}{2}N^2 - \frac{g_{12} -g}{c^2}   B^{(1)} B^{(2)}  - V_{\mathrm{tr}} N \\
&\left. + \sum_{l=1,2}\left( -h\textbf{A}^{(l)} \cdot \textbf{J}^{(l)} + \frac{cm}{2} \frac{|\vec{E}^{(l)}|^2}{B^{ (l)}}  -\frac{\hbar^2}{8cm}  \frac{ |\vec{\nabla} B^{(l)} |^2 }{B^{(l)}} \right)   \right],
\end{split}
\end{equation}
where $N =   \left(\frac{B^{(1)}}{c} +  \frac{B^{(2)}}{c} \right)$. We note that the $\vec{E}^{(l)}$, $B^{(l)}$ and $\textbf{J}^{(l)}$ fields are essentially a short-hand notation for combinations of the underlying dynamical degrees of freedom $\textbf{A}^{(l)}$ and $\vec{R}^{(l)}_{m_l}$. Importantly,  $\zeta$ does not enter as a dynamical (i.e. integration) variable in the partition function~\eqref{eqn:Z} because we have integrated out the phase variables in Eq.~\eqref{eqn:conteq} and replaced them with the classical action, an approximation that is exact only for quadratic Lagrangians. This means that while the variational equations of motion for $\vec{E}^{(l)}$ and $B^{(l)}$ will depend on $\zeta$, as a result of our approximation $\zeta$ becomes an effective parameter in Eq.~\eqref{eqn:SOC_action}. In other words, we do not vary $\zeta$ with respect to the gauge fields $\textbf{A}^{(l)}$ anymore, but instead these equations of motion must be solved in conjunction with Eqs.~\eqref{eqn:conteq} and~\eqref{eqn:EB}~\footnote{E.g. consider the exact integral $\int \mathcal{D}S \mathcal{D}n \exp \left[  \frac{\rmi}{\hbar} \int \d^3 x \left(-  \frac{\hbar^2 }{2m} n |\vec{\nabla}S|^2    \right) \right] = \int \mathcal{D}S \mathcal{D}n \mathcal{D}\vec{p} \exp \left[  \frac{\rmi}{\hbar} \int \d^3 x \left(\vec{\nabla}S \cdot \vec{p} +  \frac{m \vec{p}^2}{2 \hbar^2 n}    \right) \right]$. Replacing the functional integral over $\vec{p}$ with the classical action is now exact. Holding $\vec{p}^{\mathrm{(cl)}}$ as a parameter that is not varied anymore with respect to $S$ and $n$, but solving the resulting equations of motion for $S$ and $n$ in conjunction with the `constraint' $\vec{p}^{\mathrm{(cl)}} = -\frac{\hbar^2}{m}  n \vec{\nabla} S$ reproduces identically the equations of motion obtained directly from the original path integral. In our case substituting the classical action is not exact, but the procedure remains the same.}.

\textit{Dynamical equations of the $U(1)$ gauge theory}: In effect, we have performed a variable transformation from the Gross-Pitaevskii variables $n_l$ and $S_l$ to the gauge fields $\textbf{A}^{(l)}$, where vortices appear as quantized point charges. However, separating the topological defects as localised point-like degrees of freedom from the other degrees of freedom provides the most natural and physically transparent coordinates to study vortices. 

The first dynamical equation for the dual $U(1)$ theory we obtain is the Maxwell-Faraday law $\frac{1}{c} \partial_t B^{(l)} = - \vec{\nabla} \times \vec{E}^{(l)}$. Using the definitions~\eqref{eqn:EB} of the $\vec{E}^{(l)}$ and $B^{(l)}$ fields we find the continuity equation~\eqref{eqn:conteq}.

Having obtained the partition function, we are now in a position to derive equations of motion by varying the action~\eqref{eqn:SOC_action}. Variation of the action with respect to $a_0^{(l)}$ gives an analogue of Gauss' law:
\begin{equation}
\label{eqn:eom1_gauss}
\begin{split}
\vec{\nabla} \cdot \left( \frac{\vec{E}^{(l)} }{ B^{(l)}} \right) &=  \frac{h}{cm} J_0^{(l)},
\end{split}
\end{equation}
where the time-like component of the vortex 3-current $J_0^{(l)}$ is defined in Eq.~\eqref{eqn:vortexcurrent}. Equation~\eqref{eqn:eom1_gauss} encompasses the quantization of the circulation of vortices in the context of a compressible SO coupled BEC. In particular, Eq.~\eqref{eqn:eom1_gauss} is a transparent way to understand how vortices map to the charges in the $U(1)$ gauge theory. Let us assume that we have a single vortex at the origin, and we are in the incompressible limit. Then, the vortex acts as a delta function source (or sink) of the $\vec{E}^{(l)}/B^{(l)}$ field, which is radial in the vicinity of the vortex core. Noting that the operator $\hat{\epsilon}$ rotates by $-\pi/2$ in the $xy$-plane, we see that a radial $\vec{E}^{(l)}/B^{(l)}$ field corresponds to circular superfluid flow. In 2D the field decays as $1/r$ leading to the usual logarithmic potential between two vortices, due to the vortices alone, but in general Eq.~\eqref{eqn:eom1_gauss} must be solved in conjunction with the other equations that we derive.

Variation of the action~\eqref{eqn:SOC_action} with respect to $\vec{a}^{(l)}$ gives the Euler-Amp\`ere equation:
\begin{equation}
\label{eqn:eom3_Euler}
\begin{split}
0 &= h \vec{J}^{(l)}  + m\frac{\partial}{\partial t} \left(\frac{\vec{E}^{(l)}}{B^{(l)}} \right) \\
& -  \hat{\epsilon}\vec{\nabla} \left[ \frac{V_{\mathrm{tr}}}{c} + \frac{g}{c^2}\left(B^{(1)} + B^{(2)} \right) + \frac{g_{12} -g}{c^2}B^{(l^\prime)}  \right. \\ 
& \left. +\frac{cm}{2} \frac{|\vec{E}^{(l)}|^2 }{\left(B^{(l)}\right)^2} - \frac{\hbar^2}{8cm}    \frac{|\vec{\nabla} B^{(l)}|^2 }{\left(B^{(l)}\right)^2} +\frac{\hbar \Omega}{2c} \,  \cos{\left(\zeta\right)} \sqrt{\frac{B^{(l^\prime)}}{ B^{(l)}}} \right],
\end{split}
\end{equation}
where $l^\prime = 1,2$ if $l = 2,1$.  Thus, the equations we need to solve self-consistently for the 8 degrees of freedom $\vec{E}^{(l)}$,  $B^{(l)}$, $\zeta$ and  the total phase are (1) the continuity equation~\eqref{eqn:conteq}, which we express as the Maxwell-Faraday law $\frac{1}{c} \partial_t B^{(l)} = - \vec{\nabla} \times \vec{E}^{(l)}$ through Eqs.~\eqref{eqn:EB}; (2) the Gauss' law~\eqref{eqn:eom1_gauss}; and (3) the Euler-Amp\`ere law~\eqref{eqn:eom3_Euler}.

Variation of the action~\eqref{eqn:SOC_action} with respect to the vortex position $\vec{R}_{j}^{(l)}(t)$ gives the force $\vec{f}^{(l)}_j$ acting on the $j$th vortex in component $l$, a generalisation of the transverse vortex Magnus force for SO coupling. A detailed calculation gives
\begin{equation}
\label{eqn:eom2_Lorentz}
\begin{split}
\vec{f}^{(l)}_j &= \frac{h}{m} q_j^{(l)}\left( \vec{E}^{(l)}_j  + B^{(l)}_j \hat{\epsilon}\frac{1}{c} \partial_t \vec{R}^{(l)}_{j}\right) \\
&\qquad + \frac{\hbar\Omega}{cm}\int \d x\, \d y\, \left[ \sqrt{B^{(1)}B^{(2)}} \,\sin{\left(\zeta\right)}\frac{\partial \zeta}{\partial \vec{R}_j^{(l)}}\right],
\end{split}
\end{equation}
where the subscript in $\vec{E}^{(l)}_j$ and $B^{(l)}_j$ means that these quantities are evaluated at the vortex position, excluding the fields created by the vortex $j$ itself. The first line is the familiar transverse Magnus force, given in terms of the electrodynamic Lorentz force analogue. The second line is due to the Rabi coupling, and involves a non-local integration over the entire condensate of the Rabi energy density $V_\mathrm{n}$. 

In the limit of weak Rabi coupling and the case of a vortex pair, i.e. $\lambda_\mathrm{J} \gg d$, where $d$ is the separation between the vortices and $\lambda_\mathrm{J} =\sqrt{\hbar/(2 \Omega m)}$ is the Josephson penetration length, we can take the usual unperturbed vortex ansatz $S_l = q_j^{(l)} \mathrm{arg}\left[(x-x_j^{(l)}) + \rmi (y - y_j^{(l)})\right]$, which in an unbound uniform condensate results in an energy cost that scales as $d^2$~\cite{PhysRevA.95.023605}. This corresponds to a vortex plasma where the individual nature of the vortices is mostly preserved~\footnote{Strictly speaking there is a domain wall, but it is large with a thickness of $\sim \lambda_\mathrm{J}$.}. However, to minimise the energy especially in the limit of strong Rabi coupling $\lambda_\mathrm{J} \ll d$, the system adjusts itself such that non-zero phase is confined as a thin domain wall of thickness $\sim \lambda_\mathrm{J}$ between the vortices~\cite{PhysRevA.65.063621}, which breaks down into further vortex pairs when stretched, analogous to quark confinement~\cite{PhysRevA.93.043623}. This corresponds to a dielectric system where the charges (vortices) are confined into pairs. The transition between the vortex plasma and the dielectric phase resembles the BKT phase transition. Thus, it follows that vortices are only allowed as pairs that are bound together by the sine-Gordon kink tension~\cite{PhysRevA.92.063608,PhysRevLett.93.250406,PhysRevLett.111.170401}. Our result~\eqref{eqn:eom2_Lorentz} explains these features in terms of the vortex experiencing three contributions: in addition to the effect by the other vortices and the SO coupling parameter $k_0$ (the $\vec{E}^{(l)}_j$ field), density inhomogeneity (the $B^{(l)}_j$ field), the second line in Eq.~\eqref{eqn:eom2_Lorentz} represents the kink tension.

As an example, we now derive an exact result for the effect of $k_0$ alone on vortices by focussing on the incompressible limit~\footnote{I.e. (i) $|g_{12}| \ll |g|$ so that the energy cost of the corresponding terms in the Hamiltonian~\eqref{eqn:E} is such as to favour $n_1 \approx n_2$; and (ii) $|\Omega| \ll gn_\infty$ which means that phase fluctuations are energetically cheaper than density fluctuations. Suppressing all density fluctuations we write $n_1 = n_2 = n_\infty$.} where we also take $\Omega = 0$ lifting our approximation. Then $B^{(l)} = cn_\infty$, and $\vec{E}^{(l)} = (\hbar n_\infty/m)\hat{\epsilon} \vec{\nabla}\tilde{S}_l = -(\hbar n_\infty/m)(\vec{\nabla}\tilde{\chi}_l )$ where $\tilde{\chi}_l = \chi \pm k_0 y$ is the stream function of component $l$ with $+$ $(-)$ corresponding to $l = 2$ ($l=1$). Equation~\eqref{eqn:conteq} tells $\tilde{S}_l$ is harmonic, and Eq.~\eqref{eqn:eom1_gauss} relates the laplacian of $\chi$ with the vorticity. Unlike for the new variables, Eqs.~\eqref{eqn:eom3_Euler} and~\eqref{eqn:eom1_gauss} are valid in terms of the phase only where $\textbf{J}^{(l)} = \textbf{0}$, an advantage that carries over from the Popov formalism for the scalar case~\cite{Klein2014195}. In this limit, the force due to $k_0$ on a single vortex of circulation $q$ follows from Eq.~\eqref{eqn:eom2_Lorentz}:
\begin{equation}
\vec{f}^{(l,k_0)} = \pm \frac{2\pi q  \hbar^2 n_\infty}{m^2} k_0 \hat{\textbf{y}},
\end{equation}
where $+$ $(-)$ corresponds to $l = 2$ ($l=1$). This is nothing more but the Magnus force that the vortex experiences as a result of the homogeneous background flow along $x$ induced by the SO coupling. Once the vortex moves it will additionally experience the usual Magnus force perpendicular to its motion resulting in complicated dynamics.

The formalism presented here can also shed light on the sine-Gordon domain wall between the vortices in terms of the instanton effect~\cite{1987gauge, A&S}. Let us again consider the incompressible limit so that Eq.~\eqref{eqn:conteq} reduces to the sine-Gordon model with the action $\mathcal{S}_{\mathrm{sG}} = \int \d^3 x \left[\frac{1}{2} \left(\partial_\mu  \zeta \right)^2 - \eta  \cos{\left( \zeta \right)} \right] $, where  $\eta = \Omega m/\hbar$ for the relative phase, and the free wave equation for the total phase. Expansion of $\mathcal{S}_{\mathrm{sG}}$ in powers of $\eta$ can in general be mapped to instanton solutions of a (1+2)-dimensional Wick-rotated $U(1)$ gauge theory $\mathcal{L}_{U(1)} \propto \vec{E}^2 + B^2$~\cite{Wen2004}, which is not dissimilar from Eq.~\eqref{eqn:SOC_action} in the incompressible limit where we take $\Omega/(gn_\infty) \ll 1$ suppressing our approximation. The instanton effect corresponds to a set of magnetic charges with Coulomb interactions~\cite{1987gauge}, giving the gauge boson a finite energy gap (mass) and causing confinement of the $U(1)$ gauge charges in 1+2 dimensions~\cite{Wen2004,1987gauge}, which is consistent with compact Abelian gauge theories having been shown to exhibit confinement in 1+1 and 1+2 dimensions~\cite{2001afpp.book.1945S,1987gauge}. Conceptually, the Rabi coupling induces tunneling transitions between the spin components, which on the $U(1)$ gauge theory side of the duality are represented by the instanton gas. The conversion rate of the spin states is the highest near the domain wall, and essentially non-existent far away from it, changing sign across the domain wall~\cite{PhysRevA.65.063621}. In this sense, the domain wall can also be viewed as a Josephson `vortex line'. The action $\mathcal{S}_{\mathrm{sG}}$ possesses $Z_1$ symmetry ($\zeta \to \zeta + 2\pi)$, but not the full $U(1)$ phase symmetry of $\zeta \to \zeta + f$, where $f$ is an arbitrary constant - if it did, then the flux-non-conserving instanton effect would not be possible on the $U(1)$ gauge theory side. Only the total phase sector has the full $U(1)$ symmetry. In this sense the Rabi coupling `locks' the relative phase.

\textit{Conclusions:} Equations~\eqref{eqn:conteq},~\eqref{eqn:eom3_Euler},~\eqref{eqn:eom1_gauss} and~\eqref{eqn:eom2_Lorentz} form the key result of this work. Our formalism represents a hydrodynamical description of the SO coupled BEC that explicitly includes vortex contributions. The non-linear electrodynamics formulation that we have derived here does not have Galilean invariance because the Hamiltonian~\eqref{eqn:h0} itself breaks parity, time reversal and Galilean invariance. Rather remarkably, however, $h_0$ is translationally invariant in the absence of trapping $V_\mathrm{tr}$. Therefore, one should expect a vortex mass to be generated as a result of the broken translational symmetry due to $V_\mathrm{tr}$, as has been shown in detail, for example, for a vortex confined between two parallel walls~\cite{1367-2630-19-2-023029}, however, we leave this for a future investigation.

I would like to thank Joachim Brand for fruitful discussions and important feedback on the manuscript, and Xiaoquan Yu, Antonio Mu\~noz Mateo and Avraham Klein for fruitful discussions.

\bibliographystyle{apsrev4-1}
\bibliography{Fermi_Gas,Solitons,references}

\end{document}